\documentclass[sn-mathphys-num]{sn-jnl}
\usepackage{amsfonts}
\usepackage{graphicx}
\usepackage{amsmath}
\usepackage{amssymb}
\usepackage{mathtools}
\usepackage{braket}

\title{Construction of Metaplectic Representations of \texorpdfstring{$SL_2(\mathbb{Z}_{2^n})$}{SL2Z2n} and Twisted Magnetic Translations}
\author[1,2]{E. Floratos}

\author*[1]{K. Manolas}
\email{kmanolas@phys.uoa.gr}

\author[1]{I. Tsohanjis}
\affil[1]{Faculty of Physics, Department of Nuclear and Particle Physics, National and Kapodistrian University of Athens, University Campus, Ilisia, 15771 Athens, Greece}
\affil[2]{Academy of Athens, Panipistimiou 28, 10679 Athens, Greece}


\abstract{Unitary metaplectic representations of the group \texorpdfstring{$SL_2(\mathbb{Z}_{2^n})$}{SL2Z2n} are necessary to describe the time evolution of $2^n$-dimensional quantum systems, such as systems involving $n$ qubits. It is shown that in order for the metaplectic property to be fulfilled, an increase in the dimensionality of the involved $n$-qubit Hilbert spaces, from \texorpdfstring{$2^n$}{2n} to \texorpdfstring{$2^{2n}$}{22n}, is necessary. Thus we construct the general matrix form of such representations based on the magnetic translations of the diagonal subgroup \texorpdfstring{$HW_{2^n} \otimes HW_{2^n}$}{HW2nHW2n}.  Comparisson with other approaches on this problem of the literature are discussed.}

\begin{document}

\maketitle

\newpage
\tableofcontents

\section{Introduction}
The last decade has seen an explosive rise in interest surrounding Quantum Information and Quantum Computation, with a major focus on both constructing Quantum Algorithms as well as constructing physical implementations of Quantum Computers. 

As the physical implementations of these quantum systems require isolation from the environment, restricting the energy transitions to transitions between a fixed finite number of them, the quantum mechanical problem reduces to the study of finite dimensional Hilbert spaces. Quantum Mechanics on a finite lattice was first introduced by Weyl \cite{Weyl}, and further elaborated upon in \cite{Schwinger, HANNAY1980, Itzykson1986, FLORATOS1989, FLORATOS1997, Vourdas_2004}. There have been various applications of finite dimensional Hilbert spaces in the literature, among which are applications to the Quantum Hall effect \cite{Zak1989, Rammal1990}, studies in regards to the unification of gravity and quantum mechanics \cite{Athanasiu_1996, Axenides2014, Banks2018, Bao2017, singh2020}, as well as connections to quantum chaos \cite{berry1987}.

On the other hand, despite the growing interest, construction of quantum algorithms is still mainly done empirically, without a solid theoretical structure to be based upon.
Since a Quantum Algorithm can be viewed as an $N$-dimensional unitary operator, representing an evolution operator of the physical system, the problem can be reformulated as one of constructing an $N \times N$ matrix obeying specific criteria.

One approach for obtaining such matrices is through the construction of exact matrix representations of the group $SL_2(\mathbb{Z}_N)$. The group $SL_2(\mathbb{Z}_N)$ is the group of automorphisms of the discrete Heisenberg-Weyl group $HW_N$, whose inequivalent irreducible unitary representations have all been determined in \cite{Floratos2025} (see also \cite{Schulte2004, GUREVICH2012}). The matrix representations of $SL_2(\mathbb{Z}_N)$ have been extensively studied in the particular case of $N$ being an odd prime \cite{Itzykson1986, ATHANASIU1994, Eholzer1995, Athanasiu_1998}.

In this work we attempt to connect Finite Quantum Mechanics (FQM) and Quantum Information by constructing exact unitary representations for $SL_2(\mathbb{Z}_{2^n})$, which correspond to evolution operators of $n$-qubit quantum systems, and consequently are of great practical interest for Quantum Computation applications through the construction of the corresponding quantum circuits. The construction of exact representations for $N$ even poses specific challenges, stemming from the use of modular algebra, which do not exist for the case of $N$ being odd, for which some quantum circuits have already been proposed \cite{floratos2025Qudits}.

The paper is organised as follows:

In Section 2 we give a brief overview of the Heisenberg-Weyl group $HW_N$ and summarise the results of Floratos et al \cite{Floratos2025}, where the representations of $HW_{2^n}$ have been categorised. 

In Section 3 we describe the automorphism group of $HW_N$, $SL_2(\mathbb{Z}_N)$, and introduce the Magnetic Translations for the case of $N=odd$, in order to introduce the main notions, of a particle moving on a discrete toroidal phase space $T_2[N]$, of FQM.

In section 4 we define the Magnetic Translations for $N = 2^n$ on the Twisted double torus $T_2[2^n] \otimes T_2[2^n]$, terming them Twisted Magnetic Translations, and provide their general matrix elements.

In Section 5 we first provide the explicit matrix forms for the representations of the generators $S$ and $T$ of the group $SL_2(\mathbb{Z}_N)$, and show that they satisfy the metaplectic property. We then give the main result of this work, which is the general matrix form of the representation of any element $A \in SL_2(\mathbb{Z}_N)$. 

In Section 6 briefly overview of the results of Nobs et al \cite{Nobs1976, Nobs1976_2} and Eholzer \cite{Eholzer1995}, which we leverage to prove the exactness of our representation. Then we comparatively review the results of Feichtinger et al \cite{Feichtinger2006}, who have also constructed operators of $SL_2(\mathbb{Z}_N)$ for any $N$, and demonstrate the necessity of our construction for physical applications.

Finally, in Section 7 we conclude and discuss open issues and potential applications.

\section{The Heisenberg-Weyl Group and a set of its representations}
The discrete and finite Heisenberg-Weyl group $HW_{N}$, of order $%
|HW_{N}|=N^{3}$, can be defined as the set of elements $g(m,s,r) \equiv z^{m}x^{s}y^{r}$ where $%
m,r,s=0,1,\dots ,N-1$ and $x,y,z$ are subject to the relations 
\begin{equation}
x^{N}=y^{N}=z^{N}=e,\ yx=zxy,  \label{1}
\end{equation}%
\begin{equation}
zx=xz,\ zy=yz  \label{2}
\end{equation}%
with $e$ being the identity element of the group. Within the framework of
FQM, the generators $x$ and $y$ of \eqref{1} correspond to the position and
translation operators respectively, of a fictitious particle hopping on a discrete circle of
$N$ equidistant points whose corresponding phase space is the discrete
torus $T_{2}[N]=\mathbb{Z}_{N}\times \mathbb{Z}_{N}$. 

For physical implementations of the Heisenberg group $HW_{2^n}$ on the
toroidal phase space $T_2[2^n]$, the complete set of irreducible representations have been found and classified in \cite{Floratos2025}. In this work we shall only consider the inequivalent
faithful unitary irreducible matrix representations $\Gamma ^{p}$ of $HW_{2^{n}}$, given by
\begin{equation}
\Gamma ^{p}(z^{m}x^{r}y^{s})_{k,j} \equiv \Gamma _{m,r,s}^{p} =e^{\frac{2\pi i}{2^{n}}pn}e^{\frac{2\pi i%
}{2^{n}}pkr}\delta _{k+s-j,0}  \label{3}
\end{equation}
where $k,j,m,r,s=0,1,\dots ,2^{n}-1$ and $p=1,3,\dots ,2^{n}-1$. Any two
irreducible representations obey the following commutation relation 
\begin{equation}
\left[ \Gamma _{m,r,s}^{p},\Gamma _{m^{\prime },r^{\prime },s^{\prime }}^{p}%
\right] =(\omega _{n}^{pr^{\prime }s}-\omega _{n}^{prs^{\prime }})\Gamma
_{m+m^{\prime },r+r^{\prime },s+s^{\prime }}^{p}\ ,\quad m,r,s=0,1,\dots
,2^{n}-1  \label{4}
\end{equation}%
where $\omega _{n}=e^{\frac{2\pi i}{2^{n}}}$ and we have defined $\Gamma
^{p}(z^{m}x^{r}y^{s})\equiv \Gamma _{m,r,s}^{p}$ for notational simplicity.

The carrier space of the above irreps $\Gamma^p$ is given by
\begin{equation}  \label{5}
V^p = \{span\ket{j}\}, j=0,1, \dots, 2^n - 1
\end{equation}
where the canonical basis $\ket{j} = (e_j)_k = \delta_{k,j}$, obeying $%
\braket{k|j} = \delta_{k,j}$, is used.

The generators $x$ and $y$ of \eqref{1} correspond to the position
operator in the diagonal form and the translation (momentum) operator of a fictitious
particle moving on a discrete $2^n$ equidistant point circle, respectively.
The corresponding phase space is the discrete torus $T_2[2^n] = \mathbb{Z}%
_{2^n} \times \mathbb{Z}_{2^n}$. Given the action of the elements $z,x,y$ on
the above basis 
\begin{equation}  \label{grp_action}
\begin{split}
z \ket{j} = \omega^p_n \delta_{k,j} \ket{k},& \quad x \ket{j} =
\omega^{pk}_n \delta_{k,j} \ket{k} \\
y \ket{0} = \ket{2^n - 1},& \quad y \ket{j} = \delta_{k+1, j} \ket{k}
\end{split}%
\end{equation}
for $j = 1, \dots, 2^n - 1$, the $2^n \times 2^n$ matrix forms of $x, y$ and 
$z$ are 
\begin{align}
z_p \equiv \Gamma^p_{(1,0,0)} &= \omega_n^p I_{2^n}  \label{6} \\
Q_p \equiv \Gamma^p_{(0,1,0)} &= 
\begin{bmatrix}
1 & 0 & \dots & 0 \\ 
0 & \omega^p_n & \dots & 0 \\ 
\vdots & \ddots & \ddots & \vdots \\ 
0 & \dots & 0 & \omega_n^{p (2^n - 1)}%
\end{bmatrix}
\label{7} \\
P^{-1}_p \equiv \Gamma^p_{(0,0,1)} &= 
\begin{bmatrix}
0 & 1 & 0 & 0 & \dots & 0 \\ 
0 & 0 & 1 & 0 & \dots & 0 \\ 
\vdots & \vdots & \ddots & \ddots & \ddots & \vdots \\ 
0 & 0 & \dots & \ddots & 0 & 1 \\ 
1 & 0 & 0 & \dots & 0 & 0%
\end{bmatrix}
\label{8}
\end{align}
from which it can be shown that 
\begin{equation}  \label{9}
P^{-1}_p Q_p = \omega_n^p Q_p P^{-1}_p
\end{equation}
through whom the expected $2^n$ periodicity can be derived: 
\begin{equation}  \label{10}
Q_p^{2^n} = P_p^{2^n} = z_p^{2^n} = I_{2^n}
\end{equation}

Using \eqref{7}, the normalised eigenvectors and corresponding eigenvalues
of $P_p$, are respectively given by: 
\begin{equation}  \label{11}
\begin{split}
\ket{\psi_k} &= \frac{1}{\sqrt{2^n}} \left[ 1, \omega_n^k, \omega_n^{2k},
\dots, \omega_n^{(2^n - 1)k} \right]^T \\
\lambda_k &= \omega_n^k \ , \quad k = 0, 1, \dots, 2^n - 1
\end{split}%
\end{equation}
Then the diagonilising matrix $F_n$ of $P^{-1}_p$, with matrix elements 
\begin{equation}  \label{12}
(F_n)_{kj} = \frac{1}{\sqrt{2^n}}\omega_n^{kj}
\end{equation}
and the matrix form 
\begin{equation}  \label{13}
F_n = \frac{1}{\sqrt{2^n}} 
\begin{bmatrix}
1 & 1 & 1 & \dots & 1 \\ 
1 & \omega_n & \omega_n^2 & \dots & \omega_n^{(2^n - 1)} \\ 
1 & \omega_n^2 & \omega_n^4 & \dots & \omega_n^{2(2^n - 1)} \\ 
\vdots & \vdots & \vdots & \dots & \vdots \\ 
1 & \omega_n^{(2^n - 1)} & \omega_n^{2(2^n - 1)} & \dots & \omega_n^{(2^n -
1)(2^n - 1)}%
\end{bmatrix}%
\end{equation}
is the standard Finite Fourier Transform for n qubits, representing the
rotation by $\pi / 2$ degrees i.e. $F_n^4= I_p$ in the discrete phase space.
Finally, by making use of \eqref{7}, \eqref{8} and \eqref{13}, it can be
seen that: 
\begin{equation}  \label{14}
F_p (P_p)^p F_p^{-1} = Q_p \ , \quad p = 1, 3, \dots, 2^n - 1
\end{equation}

\section{The group \texorpdfstring{$SL_2(\mathbb{Z}_N)$}{SL2ZN} as the automorphism group of \texorpdfstring{$HW_{N}$}{HWN}} \label{AutGrp}
The group of automorphisms of $HW_{N}$, $SL_{2}(\mathbb{Z}_{N})$, is defined as the set of $2\times 2$ matrices $A=\left[ 
\begin{array}{cc}
a & b \\ 
c & d%
\end{array}%
\right] $ with $a,b,c,d\in \mathbb{Z}_{N}$ and $\det A=1modN$, that preserve the symplectic inner product of $T_2[N]$. Specifically, an element $A \in SL_2(\mathbb{Z}_N)$ acts on an element $g(m,r,s) \in HW_N$ as $g(m,r,s) \xrightarrow{A} g(m, ar + cs, br +ds)$, that is $x = (r,s) \rightarrow  (r,s)A =(r', s') = x'$. It can be seen that the symplectic inner product defined as $\{ x, x' \} = x \epsilon (x')^T$, where $\epsilon=\left[ 
\begin{array}{cc}
0 & -1 \\ 
1 & 0%
\end{array}%
\right]$ is the symplectic form, is preserved. Indeed
\begin{equation} \label{SympProd}
    x \epsilon (x')^T \xrightarrow{A} (xA) \epsilon (x'A)^T = x A \epsilon A^T (x')^T = x \epsilon (x')^T
\end{equation}
The finite automorphism group $SL_2(\mathbb{Z}_N)$ is generated by the elements 
\begin{equation} \label{gen}
S=\left[ 
\begin{array}{cc}
0 & -1 \\ 
1 & 0%
\end{array}%
\right] ,\quad T=\left[ 
\begin{array}{cc}
1 & 1 \\ 
0 & 1%
\end{array}%
\right] 
\end{equation}%
as
\begin{equation} \label{SL2ElemDecomp}
A=\left[ 
\begin{array}{cc}
1 & 0 \\ 
c a^{-1} & 0%
\end{array}%
\right] \left[ 
\begin{array}{cc}
a & 0 \\ 
0 & a^{-1}%
\end{array}%
\right] \left[ 
\begin{array}{cc}
1 & a^{-1}b \\ 
0 & 1%
\end{array}%
\right] = T_L D T_R
\end{equation}
where $T_L$ and $T_R$ are the left and right translations respectively, and the dilatations D are given by 
\begin{equation}\label{D}
D(a)=\left[ 
\begin{array}{cc}
a & 0 \\ 
0 & a^{-1}%
\end{array}%
\right] =T^{-a}ST^{-a^{-1}}S^{-1}T^{-a}S^{-1}
\end{equation}%
and $a,b\in \mathbb{Z}_{N}^{\ast }=\{x\in \mathbb{Z}_{N}:gdc(x,N)=1\}$. The above generator elements obey the relations
\begin{equation}\label{DSTrel}
\begin{split}
T^{N}=I,\quad S^{2}& =D(-1) \\
D(a)D(b)=D(ab),\quad D(a)T& =T^{a^{2}}D(a),\quad SD(a)=D(a^{-1})S
\end{split}%
\end{equation}%
and the periodicity relations $S^{4}=I_{N}$ and $(ST)^{6}=I_{N}$.

In Finite Quantum Mechanics \cite{Itzykson1986, ATHANASIU1994, Vourdas_2004} we replace the points $x(r,s)$ of the toroidal phase space $T_2[N]$ by unitary $N \times N$ matrices coming from the faithful representation of the Heisenberg group
\begin{equation}  \label{jrsDefOdd}
J_{r,s} = \omega^{\frac{rs}{2}}P^r_N Q^s_N \quad r, s = 0, 1, \dots, N - 1, \ \ \omega = \exp^{\frac{2 \pi i}{N}}
\end{equation}
with $Q$ and $P$ representing the position and momentum operators of a particle moving on the discrete phase space $T_2[N]$. These matrices are called Magnetic Translations. The operators $Q$ and $P$ satisfy the fundamental Heisenberg-Weyl commutation relation $QP = \omega PQ$ which is the exponential form of the standard Heisenberg commutation relations of Quantum Mechanics. This is the faithful irrep of the finite $HW_N$ group with $p=1$ (see \eqref{6}, \eqref{7}, \eqref{8}) with matrix elements
\begin{equation}\label{QPmat_elem}
\begin{split}
(Q_{N})_{kj} &=\omega^{k}\delta _{k,j}, \\
(P_{N})_{kj} &=\delta_{k-1,j}
\end{split}
\end{equation}
where $k,j=0,1,\dots , N-1$.

The so called magnetic translations $J_{r,s}$ thus defined, provide a projective unitary
representation of the translation group on $T_{N}$ i.e. 
\begin{equation}\label{JtimesJ}
J_{r,s}J_{r^{\prime },s^{\prime }}=\omega ^{(r^{\prime }s-rs^{\prime
})/2}J_{r+r^{\prime },s+s^{\prime }}
\end{equation}
The automorphism group $SL_2(\mathbb{Z}_N)$ has the corresponding unitary representation which preservers the algebra for the magnetic translations \eqref{JtimesJ}. Thus for every $A \in SL_2(\mathbb{Z}_N)$ we correspond a unitary operator $U(A)$ with the property:
\begin{equation}\label{GeneralMetaplectic}
U^{-1}(A)J_{r,s}U(A)=J_{(r,s)A}
\end{equation}
This property is called the metaplectic property, meaning that $U(A)$ realises the action of the group element $A$ on the phase space but now on the quantum phase space consisting of the magnetic translations $J_{r,s}$.

The factor of $\frac{1}{2}$ in the cocycle \eqref{JtimesJ} is an immediate consequence
of demand of unitarity of $J_{r,s}$ i.e. $\left( J_{r,s}\right) ^{\dagger
}=J_{r,s}^{-1}=J_{-r,-s}$. Moreover it can be easily shown that  
\begin{equation*}
J_{r,s}J_{r^{\prime },s^{\prime }}=\omega ^{sr^{\prime }-s^{\prime
}r}J_{r^{\prime },s^{\prime }}J_{r,s}
\end{equation*}%
As a consequence of \eqref{JtimesJ} we obtain that $(J_{r,s})^{k}=J_{kr,ks}$ for $%
k=0,1,...,N-1$, which results in the periodic boundary condition on the
discrete $T_{2}[N]$, $(J_{r,s})^{N}=I$.

The matrix elements of the Weil representations of a general element $A=\left[ 
\begin{array}{cc}
a & b \\ 
c & d%
\end{array}%
\right]$, when $N$ is a prime integer, conforming to the definition of $J_{r, s}$ and satisfying the metaplectic property, \eqref{GeneralMetaplectic}, are given \cite{ATHANASIU1994} in the generic case by 
\begin{equation}\label{GenMatElemOdd}
    U(A)_{l,m} = \frac{1}{\sqrt{N}}(-2 c\ |\ N)\left\{ 
\begin{array}{c}
1\\ 
-i
\end{array}
\right\} \omega^{-(al^2 + dm^2 - 2lm)/ 2c}
\end{equation}
where the bracket equals $1$ if $N=4k+1$ or $-i$ if $N=4k-1$, and $(-2 c\ |\ N)$ is $+1$ if $-2c$ is the square of an integer $modN$ and $-1$ if not.
Specifically, for the representations of the generator elements $S$ and $D$, of $SL_2(\mathbb{Z}_N)$, the matrix elements are given by 
\begin{equation}\label{SmatelemOdd}
    U(S)_{l, m} = (-1)^{N} \frac{i^t}{\sqrt{N}} \omega^{lm}
\end{equation}
where $t = 0, 1$, depending on whether $N = 4 k + 1$ or $N = 4 k - 1$, respectively, and
\begin{equation}
    U(D)_{l,m} = \sigma(1) \sigma(2-a-a^{-1}) \delta_{l, am}
\end{equation}
where:
\begin{equation}
    \sigma(r) = (r \ | \ N) \left\{ 
\begin{array}{c}
1, \quad N=4k+1\\ 
-i, \quad N=4k-1
\end{array}
\right.
\end{equation}
It should be noted that the irreps of $SL_2(\mathbb{Z}_N)$ have been explicitly classified using Weil representations in \cite{Nobs1976}, \cite{Nobs1976_2}, \cite{Eholzer1995}. In the next section we shall consider the case $N=2^{n}$, being of particular interest in quantum information and
computation.

\section{Twisted \texorpdfstring{$HW_{N}$}{HWN} and Magnetic Translations for \texorpdfstring{$N=2^n$}{N2n}}
Turning now to our main case of interest of $N=2^n$, it is easily seen that an analogue
definition of $J_{r,s}$ as in \eqref{jrsDefOdd} fails to exist as division
by 2 $modN$ is undefined. A possible solution to this problem is an extension of the phase space to
the double toroidal phase space $T_2[2^n] \otimes T_2[2^n]$ with the
appropriate definition of its translation group in the tensor product group $%
HW_{2^n} \otimes HW_{2^n}$.

We begin by considering the diagonal subgroup $HW_{2^n}^d$, consisting of
all pairs $(g, g), g \in HW_{2^n}$. The tensor product of unitary irreps of $%
HW_{2^n}$ gives the unitary irreps of $HW_{2^n} \otimes HW_{2^n}$. On $%
HW_{2^n}^d$ these representations are generally reducible and equivalent to
some tensor product of unitary irreps of $HW_{2^n}$. To this end, let $x_d =
x_p \otimes x_p$, $y^{-1}_d = y^{-1}_p \otimes y^{-1}_p$ and $z_d = z_p \otimes z_p \in HW_{2^n}^d$,
where this time $x_d, y^{-1}_d$ are interpreted as position and momentum on the subspace $T_{2^n}^d$, respectively. Similarly to the odd case, these are also subject to the relations \eqref{1} $x_d y^{-1}_d = z_d y^{-1}_d x_d$.

Consider the set of elements $J_{r,s}$, $r, s =0,...,2^n-1$ of $%
HW_{2^n}\otimes HW_{2^n}$, defined by 
\begin{equation}  \label{jrsdef}
J_{r,s}\equiv (z^{-sr}\otimes I)x_d^s (y^{-1}_d)^r
\end{equation}
which satisfy 
\begin{equation*}
J_{r,s}J_{r^{\prime },s^{\prime }}=(z^{sr^{\prime }} \otimes z^{-s^{\prime
}r})J_{r+r^{\prime }, s+s^{\prime }}
\end{equation*}
Note that the set of elements $J_{r,s}$ thus defined do not all belong to $%
HW_{2^n}^d$ and moreover, form a central extension of the translation
group of the torus $T_{2^n}^d$ which we shall call $MT_{2^n}$ (Magnetic
Translations of the torus). Using the previously constructed faithful irreps of $%
HW_{2^n}$ \eqref{4}, and setting $Q_p=\Gamma^p(x),\ P_p=\Gamma^p(y^{-1}),\ z_p=\Gamma^p(z)$ the elements $J_{r,s}$ are represented by the $2^{2n}$-dimensional matrices $%
J_{r,s}^p$ 
\begin{equation}  \label{jrsrep}
J_{r,s}^p = z_p^{-rs} Q_p^s P_p^r \otimes Q_p^s P_p^r = \omega_n^{-psr}
Q_p^s P_p^r \otimes Q_p^s P_p^r
\end{equation}
The $2^{2n}$-dimensional carrier space of the above representation is simply
taken to be the tensor product of the carrier space $V^p$ of \eqref{5} with
itself 
\begin{equation}  \label{repv}
V_{2n} \equiv V^p\otimes V^p=span\{\ket{j_1, j_2},\ j_1, \ j_2 =0,1,...,2^n
- 1\}
\end{equation}
this time obeying the orthogonality relation $\braket{k_1, k_2 | j_1, j_2} =
\delta_{k_1, j_1} \delta_{k_2, j_2}$.

The matrix elements of $J_{r,s}^p$ are given by
\begin{equation}  \label{matrixv}
(J_{r,s}^p)_{(k_1,k_2),(j_1,j_2)} \equiv (J_{r,s}^p)_{2^n k_1 + k_2, 2^n j_1 + j_2} = \omega_n^{p[-sr + (k_1 + k_2)s]} \delta_{k_1 - r, j_1} \delta_{k_2-r, j_2}
\end{equation}
where $r, s = 0,...,2^n-1$ and $k_1, k_2, j_1, j_2 = 0,1,...,2^n-1$. We shall call this group defined by \eqref{matrixv} the Twisted $HW_{2^n}$ group. Using (\ref{matrixv}) it is easily shown that $J_{r,s}^p$ realize a unitary projective representation of the translation group on $T_{2^n}^d$: 
\begin{eqnarray}
J_{r,s}^p J_{r^{\prime }, s^{\prime }}^p &=& \omega_n^{p(r^{\prime }s -
s^{\prime }r)} J_{r + r^{\prime },s + s^{\prime }}^p  \label{proj} \\
(J_{r,s}^p)^{\dagger} &=& J_{-r, -s}^p  \label{uni}
\end{eqnarray}

Since $p = 1, 3, \dots, 2^n - 1$ runs through the faithful inequivalent
irreps of $HW_{2^n}$, we thus obtain all the faithful unitary (though not
necessarily irreducible) representations of the magnetic translation group $MT^d_{2^n}$. It can be easily checked that $J_{r,s}^p$ given by \eqref{jrsrep} has the periodic conditions $J_{r + 2^n,s + 2^n}^p = J_{r,s}^p$.

\section{Explicit realisations of the metaplectic representation of \texorpdfstring{$SL_2(\mathbb{Z}_{2^n})$}{SL2Z2n}}
Having constructed the twisted magnetic translations for $N=2^n$, we can proceed with the quantisation of the classical motion of the torus $T_2[2^n]\otimes T_2[2^n] $, by finding the representations of the generators of $SL_2(\mathbb{Z}_{2^n})$, $U(S)$ and $U(T)$, that fulfill \eqref{GeneralMetaplectic} for the above $J_{r,s}^p$. We adapt the ansaetze for the generators, appropriate for the Twisted $HW_{2^n}$, given by 
\begin{align}  
\left( U(S) \right)_{(k_1, k_2), (j_1, j_2)} &\equiv \left( U(S)
\right)_{2^nk_1 + k_2, 2^nj_1 + j_2} = \frac{1}{2^n} \omega_n^{p(k_1 j_2 +
k_2 j_1)} \label{Srep_elem} \\
\left( U(T) \right)_{(k_1, k_2), (j_1, j_2)} &\equiv \left( U(T)
\right)_{2^nk_1 + k_2, 2^nj_1 + j_2} = \omega_n^{-k_1 k_2} \delta_{k_1 , j_1} \delta_{k_2 , j_2} \label{Trep_elem}
\end{align}
where $k_1, k_2, j_1, j_2 = 0,1, \dots, 2^n - 1$. It is easily shown that $%
U(S)$ and $U(T)$ satisfy the algebraic relations corresponding to the classical ones given by \eqref{DSTrel} as well as the periodicity relations $U^4(S) = I_{2^{2n}}$ and $\left( U(S)U(T) \right)^6 = I_{2^{2n}}$, with both of them also satisfying the \textit{metaplectic
property} 
\begin{eqnarray}
U(S)^{-1}J_{r,s}^{p}U(S) &=&J_{(r,s)S}^{p} \label{metapl_S} \\
U(T)^{-1}J_{r,s}^{p}U(T) &=&J_{(r,s)T}^{p} \label{metapl_T}
\end{eqnarray}
while 
\begin{equation}  \label{TwistedFxy}
\begin{split}
U(S)^{-1}(Q_p \otimes I_{2^n}) U(S) &= I_{2^n} \otimes P_p \\
U(S)^{-1}(I_{2^n} \otimes Q_p)U(S) &= P_p \otimes I_{2^n}
\end{split}%
\end{equation}
corresponds to the twisted version of \eqref{14}. 
The relation between $U(S)$ and $F_n$ is given by 
\begin{equation}  \label{UFtwist}
U(S) = \tau(F_n \otimes F_n)
\end{equation}
where $F_p$ is given by \eqref{14} and $\tau$ is the twist map $\tau(\ket{a}
\otimes \ket{b}) = \ket{b} \otimes \ket{a}$. The representation of the other
generator $T$ of $SL_2(\mathbb{Z}_{2^n})$ can be expressed in terms of 
$Q_{p}$ as:
\begin{equation}  \label{TrepQ}
U(T) = \frac{1}{2^n} \sum_{s_1, s_2 = 0}^{2^n - 1} \omega_n^{p s_1 s_2}
Q_p^{s_1} \otimes Q_p^{s_1}
\end{equation}

For physical applications, a general relation giving the matrix elements of the metaplectic representation $U(A)$, for any $A\in SL_{2}(\mathbb{Z}_{2^{n}}),$ is needed. In the following, having found the representations $U(S)$ and $U(T)$, we will construct the general form of the unitary $U(A)$ for every $A \in SL_{2}(\mathbb{Z}_{2^{n}})$ on the twisted toroidal phase space and show that it is proper.

With the explicit matrix representations of the generator elements $S$ and $T$ in hand, we are now in a position to construct the general form of $U(A)$. This can be done if $A$ is written as an explicit string of products of powers of $S$ and $T$.  This problem can be overcome by considering the following decompositions for a general element $A=\left( 
\begin{array}{cc}
a & b \\ 
c & d%
\end{array}%
\right) \in SL_{2}(\mathbb{Z}_{2^{n}})$, where $a,b,c,d\in \mathbb{Z}_{2^{n}}
$ and $detA=1mod2^{n}$. To facilitate the construction we consider two cases 
$A_{o}$, $A_{e}$ depending on whether the entry $d$ in matrix $A$ above is
odd ($d\equiv d_{o}$) or even ($d\equiv d_{e}$). Using the generators $S$, $T
$ and the dilatations $D$ of \eqref{gen}, \eqref{DSTrel} and \eqref{D}, $%
A_{o}$ can be decomposed as 
\begin{eqnarray}  \label{dec1}
A_{o} &=&\left( 
\begin{array}{cc}
a & b \\ 
c & d_{o}%
\end{array}
\right) =\left( 
\begin{array}{cc}
1 & bd_{o}^{-1} \\ 
0 & 1%
\end{array}
\right) \left( 
\begin{array}{cc}
d_{o}^{-1} & 0 \\ 
0 & d_{o}%
\end{array}
\right) \left( 
\begin{array}{cc}
1 & 0 \\ 
cd_{o}^{-1} & 1%
\end{array}
\right) \notag \\
&=&T^{\frac{b}{d_{o}}}D(d_{o}^{-1})S^{-1}T^{-\frac{c}{d_{o}}}S
\end{eqnarray}
where $T$ realises left translations, and $S^{-1}T^{-\frac{c}{d_o}}S$ realises right translations.

In the case of $A_e$, where $c, b$ must be odd, the previous decomposition cannot be
applied. However, as $c$ is odd, we can decompose $A_e$ as: 
\begin{eqnarray}  \label{dec2}
A_{e} &\equiv &\left( A_{e}S^{-1}\right) S=\left( 
\begin{array}{cc}
-b & a \\ 
-d_{e} & c%
\end{array}
\right) \left( 
\begin{array}{cc}
0 & 1 \\ 
-1 & 0%
\end{array}
\right)  \notag \\
&=&T^{\frac{a}{c}}D(c^{-1})S^{-1}T^{\frac{d_{e}}{c}}S^{2}
\end{eqnarray}

We define $U(A_o)$ and $U(A_e)$ by 
\begin{align}  \label{repdec1}
U(A_{o}) &= U(T^{\frac{b}{d_o} }D(d_o^{-1})S^{-1}T^{-\frac{c}{d_o} }S)  \notag \\
&\coloneqq U(T^{\frac{b}{d_o} })U(D(d_o^{-1}))U(S^{-1})U(T^{-\frac{c}{d_o} })U(S)
\end{align}
and 
\begin{align}  \label{repdec2}
U(A_{e}) \coloneqq& U(A_{e}S^{-1})U(S)=U(T^{\frac{a}{c} }D(c^{-1})S^{-1}T^{-\frac{d_e}{c} }S)U(S)  \notag \\
=& U(T^{\frac{a}{c} })U(D(c ^{-1}))U(S^{-1})U(T^{-\frac{d_e}{c} })U(S^{2})
\end{align}
where $A_o, A_e$ are given by \eqref{dec1} and \eqref{dec2}, respectively.
Using \eqref{Srep_elem} and \eqref{Trep_elem} in \eqref{repdec1} and \eqref{repdec2},
the matrix elements of $U(A_o)$ and $U(A_e)$ are obtained as 
\begin{align}
U(A_{o})_{2^nk_{1}+k_{2},2^nj_{1}+j_{2}} &= \begin{aligned}[t]
\frac{1}{2^n} \sum_{r=0}^{2^n-1}%
&\omega ^{-\frac{b}{d_o} k_{1}k_{2}+\left( -\frac{1}{d_o } k_{2}+j_{2}\right)
r} \\ &\quad\delta _{-\frac{1}{d_o }k_{1}+\frac{c}{d_o} r+j_{1},0}
\end{aligned}\label{matrixdec1} \\
U(A_{e})_{2^nk_{1}+k_{2},2^nj_{1}+j_{2}} &= \frac{1}{2^{n}}\omega
^{-\frac{a}{c} k_{1}k_{2}}\omega ^{\frac{1}{c }\left(k_{1}j_{2}+k_{2}j_{1}\right) }\omega ^{-\frac{d_e}{c} j_{1}j_{2}}  \label{matrixdec3}
\end{align}
where $k_{1}$, $k_{2}$, $j_{1}$, $j_{2}=0,1,...,2^{n}-1$.

The sum in \eqref{matrixdec1} can be, in specific cases, be reduced to 
\begin{align}  \label{goodd}
U(A_{o})_{2^nk_{1}+k_{2},2^nj_{1}+j_{2}} = \frac{1}{2^n}\omega ^{-\left(
\frac{b}{d_o} +\frac{1}{c d_o}\right) k_{1}k_{2}-\frac{ j_{2}j_{1}}{%
\frac{c}{d_o} }}\omega ^{\frac{k_{2}j_{1}+j_{2}k_{1}}{c}} \quad \text{if} \  \frac{c}{d_o} \ \text{is odd}
\end{align}
and 
\begin{equation}  \label{gozero}
U(A_{o})_{2^nk_{1}+k_{2},2^nj_{1}+j_{2}} = \omega^{-\frac{b}{d_o} k_{1}k_{2}}
\delta_{-\frac{1}{d_o }k_{1}+j_{1},0} \delta_{-\frac{1}{d_o }
k_{2}+j_{2}} \quad \text{if \ } c = 0.
\end{equation}
It is straightforward to show that $U(A_{o}^{-1})=U(A_{o})^{-1}$ and $U(A_{e}^{-1})=U(A_{e})^{-1}$.

By evaluating all possible matrix products $U(A_{i})U(A_{j})$, $i, j=\{o,e\}$, and comparing with the corresponding matrix $U(A_{i}A_{j}),$ it can be shown that this representation is proper and faithful but not irreducible due to the non-irreducibility of the twisted $HW_{2^n}$ group. Note
that in doing so, care should be taken so that $U(A_{i}A_{j})$ should be
expressed as in \eqref{matrixdec1} or \eqref{matrixdec3} accordingly.

Although evaluating all possible matrix products is a valid approach, it requires cumbersome calculations. Instead, we will make use of results from \cite{Nobs1976_2, Eholzer1995}, a short review of which we will provide in the following section.

\section{Remarks on relevant works}
It is important to note that Weil representations and relevant irreps of Weil representations and relevant irreps of $SL_2(\mathbb{Z}_{2^n})$ have been defined and classified by Nobs et al in \cite{Nobs1976_2} and Eholzer in \cite{Eholzer1995}, using quadratic modules of $Z_{p^n}$, with $p$ being an odd prime or $2$. Our construction of unitary metaplectic representations of $SL_2(\mathbb{Z}_{2^n})$ is related to this classification, as will be demonstrated in what follows. According to \cite{Nobs1976_2, Eholzer1995} let $M$ denote a finite $\mathbb{Z}_{p^n}$-module. Let Q be a quadratic form of $M, \ Q: M \longrightarrow p^{-n} \mathbb{Z} / \mathbb{Z}$ such that
\begin{equation}\label{quadr_prop}
    Q(-x) = Q(x), \ \forall x \in M
\end{equation}
and define a $\mathbb{Z}_{p^n}$-bilinear map $B : M \times M \longrightarrow p^{-n} \mathbb{Z}/\mathbb{Z}$ as:
\begin{equation}\label{bilinear_map}
    B(x,y) := Q(x + y) - Q(x) - Q(y)
\end{equation}
The pair $(M,Q)$ is called a quadratic module of $\mathbb{Z}_{p^n}$. Then, a Weil representation $\Gamma : SL_2(\mathbb{Z}_N) \longrightarrow Aut\mathbb{C}^N$ is given once we define the action on the generators of $SL_2(\mathbb{Z}_N)$,
\begin{align}\label{Weil_act_gen}
    \left[ \Gamma(T) f \right](x) &\approxeq e^{2 \pi i Q(x)}f(x) \\
    \left[ \Gamma(S^{-1}) f \right](x) &\approxeq \frac{\alpha_Q(-1)}{|M|^{1/2}} \sum_{y \in M} \exp{2 \pi B(x,y)} f(y) \\
    \left[ \Gamma(D(a)) f \right](x) &\approxeq \alpha_Q(a) \alpha_Q(-1) f(a^{-1} x)
\end{align}
where $a \in \mathbb{Z}^*_{p^n}$, $|M|$ the order of $M$ and:
\begin{equation}\label{alpha_q}
    \alpha_Q(a) = |M|^{-1/2} \sum_{x \in M} \exp{2 \pi a i Q(x)}
\end{equation}
This defines a \textit{proper} representations of $SL_2(\mathbb{Z}_{p^n})$, provided that:
\begin{equation}\label{prop_rep_requir}
    \alpha_Q(a) \alpha_Q(b) = \alpha_Q(1) \alpha_Q(ab), \ \forall a,b \in \mathbb{Z}^*_{p^n}
\end{equation}
If this relations is not satisfied, the representation is projective. For the case of $p=2$, the quadratic modules $(M,Q)$ that define proper Weil representations are given in Theorem 4 of \cite{Nobs1976_2}. In this article we focus solely on case (1) of Theorem 2 in \cite{Eholzer1995} (see also \cite{Nobs1976_2}), where $M = \mathbb{Z}_{2^n} \times \mathbb{Z}_{2^n}$ and $Q(x)= \frac{1}{2^n} x_1 x_2 , \ x_1, x_2 \in \mathbb{Z}_{2^n}$. In this case it is easy to check that \eqref{prop_rep_requir} is satisfied, and thus the representation of $SL_2(\mathbb{Z}_{2^n})$ constructed in this work is exact (but not irreducible).

\subsection{Weyl projective representations with quadratic characters}
In the following we briefly present the work of Feichtinger et al \cite{Feichtinger2006} which also deals with the problem of constructing the metaplectic representations of $SL_2(\mathbb{Z}_N)$ for any $N$ and critically compare their results with those of ours. The notation to be followed below has been chosen so that it is consistent with ours.

Using the matrix representation of the translation and momentum operators defined as
\begin{align}
    Q_{i, j} &= e^{\frac{2 \pi i}{N} j} \delta_{i, j} \label{NeuQelem} \\
    P_{i,j} &= \delta_{i-1, j} \label{NeuPelem}
\end{align}
the \textit{time-frequency shift operator} is defined as
\begin{equation}\label{time_shift_elem}
    \pi(r,s) = P^r Q^s
\end{equation}
which corresponds to the magnetic translation elements that we have defined in the previous chapters but without the appropriate phase.
Similarly to the property of projective representation given by \eqref{JtimesJ}, here we have
\begin{equation}\label{time_shift_comp}
    \pi(\lambda) \pi(\lambda') = e^{2 \pi i \braket{\lambda | \kappa \lambda'} / N} \pi(\lambda + \lambda') , \quad \text{where}\ \kappa = \left( 
\begin{array}{cc}
0 & 0 \\ 
1 & 0
\end{array}
\right)
\end{equation}
with $\lambda = (r, s) \in \mathbb{Z}_N \times \mathbb{Z}_N$. Under this definition $\pi(\lambda)$ is clearly not unitary, but instead $\pi(\lambda)^{-1}= e^{2 \pi i \braket{\lambda | \kappa \lambda} / N} \pi(-\lambda)$ holds.
For $A = \big(\begin{smallmatrix}
  a & b\\
  c & d
\end{smallmatrix}\big) 
\in SL_2\left(\mathbb{Z}_n\right)$
the metaplectic identity of an operator $U(A)$ is 
\begin{equation}\label{Neu_meta}
    U(A)\pi(k,l)U^{-1}(A) = \psi_A(k,l) \pi(ak + bl, ck + dl)
\end{equation}
where $\psi_A$ a second degree character associated to $\sigma_A$ where
\begin{equation}\label{sigma_a}
    \sigma_A = \begin{pmatrix}
        ac & bc\\
        ad - 1 & bd
    \end{pmatrix} = A^T \kappa A - \kappa, \qquad \kappa = \begin{pmatrix}
        0 & 0\\
        1 & 0
    \end{pmatrix}
\end{equation}
and satisfies by definition the following relation
\begin{equation}\label{sec_deg_char}
    \psi_A(k + k', l + l') = \psi_A(k,l) \psi_A(k',l') exp\left(\frac{2 \pi i}{n} (k,\ l) \sigma_A \big(\begin{smallmatrix}
  k'\\
  l'
\end{smallmatrix}\big) \right)
\end{equation}
If we denote $\Psi_A$ as the class of all second degree characters $\psi_A$ associated to $\sigma_A$, then for $\psi_A, \psi'_A \in \Psi_A$ it holds that
\begin{equation}\label{sec_deg_char_class}
    \psi'_A = \chi \psi_A
\end{equation}
where $\chi = e^{\frac{2 \pi i}{n} (rk + sl)}$.

The general form of the constructed operators $U(A)$, for N both even and odd, as explicitly given by (see \textit{Theorem 1} \cite{Feichtinger2006})
\begin{equation}\label{NeuOperElem}
    \left( U \right)_{k m} = e^{\pi i \frac{N+1}{N} ([c_0a^{-1}_0]k^2 + [-\theta]m^2)} \frac{1}{N} \sum_l \omega^{l(ka_0^{-1} - m)} e^{\pi i [-a_0^{-1}b]l^2 \frac{N+1}{N}}
\end{equation}
where for $N=2^n$, $\theta = 2$ if $a =$ odd and $\theta = 1$ if $a =$ even, must satisfy the metaplectic identity \eqref{Neu_meta}.

These operators have been shown (see \textit{Lemmas 5, 7, 9} in \cite{Feichtinger2006}) to satisfy the following properties
\begin{enumerate}
\item If $U_1, U_2$ satisfy \eqref{Neu_meta} for the same $\psi_A$ then 
\begin{equation}\label{lemma5}
    U_2 = e^{i \phi} U_1
\end{equation}
\item If $U_1, U_2$ satisfy \eqref{Neu_meta} for $\psi_1$ and $\psi_2 \in \Psi_A$ accordingly, then
\begin{equation}\label{lemma7}
    U_2 = W U_1
\end{equation}
where $W$ belongs to the Heisenberg group $HW_N = \{ \tau P^rQ^s: r,s \in \mathbb{Z}_N,\ \tau \in \mathbb{C}: |\tau| = 1 \}$.
\item Given $U(A), \psi_A$ and $U(B), \psi_B$, $U(A)U(B)$ satisfies \eqref{Neu_meta} for
\begin{equation}\label{lemma9}
    \psi(k, l) = \psi_A(B \big(\begin{smallmatrix}
  k\\
  l
\end{smallmatrix}\big) ) \psi_B(k, l) \in \Psi_{AB}
\end{equation}
\end{enumerate}
which together with \textit{Proposition 1} in \cite{Feichtinger2006} show that the set
\begin{equation}\label{Neu_final_remark}
    \left\{ (A, \psi): A \in SL_2(\mathbb{Z}_N),\ \psi \in \Psi_A(\sigma_A) \right\}
\end{equation}
can be identified as a class of automorphisms of the Heisenberg group $HW_N$.

Concluding the comparison of our work with \cite{Feichtinger2006}, the authors in \cite{Feichtinger2006} obtain a set of irreducible \textit{operators} \eqref{NeuOperElem} of dimension $2^n$ which require the quadratic characters in order to be identified as a class of automorphisms of the group $HW_N$. On the other hand we obtain \textit{reducible} representations of dimension $2^{2n}$ which are exact, and consequently don not require the use of second degree characters.

Furthermore, if what we require is a general representation of $SL_2(\mathbb{Z}_N)$ that satisfies the metaplectic relation, for both $N$ even and odd, to use for physical applications, then the construction of \cite{Feichtinger2006} is unfit as it is not a representation. 

This can most easily seen through the chirp multiplication operator $R_c$, defined in \textit{Lemma 2 (iii)} of \cite{Feichtinger2006} and used in the construction of the operator \eqref{NeuOperElem}, by comparing $R_{[c_1]}R_{[c_2]}$ to $R_{[c_1 + c_2]}$ for $c_1, c_2 \in \mathbb{Z}_N$. First of all, the matrix elements of the operator are
\begin{equation}\label{NeuChirpElem}
    \left(R_{[c]}\right)_{k,l} = e^{\pi i \frac{N + 1}{N} c k^2}\delta_{k,l}
\end{equation}
through which it can be shown that in general
\begin{equation}\label{NeuRnotRep}
    \left(R_{[c_1]}R_{[c_2]}\right)_{k,l} = \theta([c_1], [c_2])^{k^2} \left(R_{[c_1 + c_2]}\right)_{k,l}
\end{equation}
where $\theta([c_1], [c_2]) = (-1)^{[c_1]mod(N+1) + [c_2]mod(N+1) - ([c_1] + [c_2])mod(N+1)}$. It can now be seen that for $N=$ odd, the general operator $U$ corresponds indeed to an exact representation of $SL_2(\mathbb{Z}_N)$, whereas for $N=$ even, \eqref{NeuRnotRep} is not a homomorphism and consequently $U$ is not a representation.

\section{Conclusion}
In this work we have provided an explicit construction of a set of reducible metaplectic matrix representations for the group $SL_2(\mathbb{Z}_{N})$, with $N=2^n$, complementing existing work in the literature, for the simpler case of $N$ odd \cite{Eholzer1995, ATHANASIU1994}. These representations are unitary and exact, meaning they can be used to describe a physical system, notably among them, an $n$-qubit quantum system. 

The requirements of unitarity and exactness come at the cost of increased computational complexity in calculating the matrix representations, with their dimension scaling faster with $n$, increasing from $\mathcal{O}(2^n)$, in the odd case, to $\mathcal{O}(2^{2n})$ here.
This increase in dimension for the case of $N = 2^n$ is also present in the general classification of irreps of $SL_2(\mathbb{Z}_{2^n})$ \cite{Nobs1976, Nobs1976_2, Eholzer1995}. 

Our work could potentially be used to construct a dynamical model mimicking the evolution of a Quantum Neural Network on a graph. This can be seen from the fact that the matrix $P$ in relation \eqref{8} corresponds to an adjacency matrix of a directed cyclic graph and the corresponding Finite Fourier Transform (FFT) which diagonilizes P and the corresponding diagonal position operator $Q$ consitute the base for Finite Quantum Mechanics on this particular cyclic graph. In a general graph the corresponding adjacency matrix plays the role of $P$ its diagonilizing matrix is the corresponding FFT of the graph. Thus one can envisage the formulation of FQM on a general graph which may be usefull for Quantum Learning on an arbitrary Graph neural Network. The role of the diagonilizing Fourier Transform of a graph is known from the applications of the Graph Convolutional Networks \cite{kipf2017}. In that context, one could also examine the role of $SL_2(\mathbb{Z}_{N})$, it being the group of automorphisms of $HW_{2^n}$. This tangent, as well as the process of finding the \textit{irreducible} representations of the twisted magnetic translations $J_{r,s}$ will be studied in the future.

Finally, note that we have limited our discussion to representations of $SL_2(\mathbb{Z}_{N})$, which have the metaplectic property for magnetic translations $J_{r,s}$ constructed using only faithful $HW_N$ irreps from \cite{Floratos2025}. A line of future research if constructed, the non-faithful representations of $SL_2(\mathbb{Z}_{N})$ could potentially be used in Quantum Machine Learning algorithms also, performing a function analogous to that of Pooling Layers in classical Convolutional Neural Networks (see Chap. 9 in \cite{Goodfellow2016}), as proposed in \cite{Nguyen2024}.

\backmatter

\bibliography{references}

\end{document}